\documentclass{article}

\usepackage{arxiv}

\usepackage[utf8]{inputenc} % allow utf-8 input
\usepackage[T1]{fontenc}    % use 8-bit T1 fonts
\usepackage{hyperref}       % hyperlinks
\usepackage{url}            % simple URL typesetting
\usepackage{booktabs}       % professional-quality tables
\usepackage{amsfonts}       % blackboard math symbols
\usepackage{nicefrac}       % compact symbols for 1/2, etc.
\usepackage{microtype}      % microtypography
\usepackage{lipsum}
\usepackage{multirow}
\usepackage{graphicx}

\title{The CNN-based Coronary Occlusion Site Localization with Effective Preprocessing Method}

\author{
    YeongHyeon Park \\
    Department of Computer and Electronic Systems Engineering\\
    Hankuk University of Foreign Studies\\
    Yongin 17035, Korea \\
    \texttt{yeonghyeon@hufs.ac.kr} \\
    \And
    Il Dong Yun \\
    Department of Computer and Electronic Systems Engineering\\
    Hankuk University of Foreign Studies\\
    Yongin 17035, Korea \\
    \texttt{yun@hufs.ac.kr} \\
    \And
    Si-Hyuck Kang \\
    Cardiovascular Center, Internal Medicine \\
    Seoul National University Bundang Hospital \\
    Seongnam 13620, Korea \\
    \texttt{eandp303@snu.ac.kr} \\
}

\begin{document}
\maketitle

\begin{abstract}
The Coronary Artery Occlusion (CAO) acutely comes to human, and it highly threats the human's life. When CAO detected, Percutaneous Coronary Intervention (PCI) should be conducted timely. Before PCI, localizing the CAO is needed firstly, because the heart is covered with various arteries. We handle the three kinds of CAO in this paper and our purpose is not only localization of CAO but also improving the localizing performance via preprocessing method. We improve localization performance from a minimum of 0.150 to a maximum of 0.372 via our noise reduction and pulse extraction based method.
\end{abstract}

% keywords can be removed

\section{Introduction}
\label{sec:introduction}

The Coronary Artery Occlusion (CAO) makes thrombus or embolism and these cause sudden blood flow shutting-down \cite{bax2012third}. The physician will conduct Percutaneous Coronary Intervention (PCI) when the patient is suffered from CAO. However, the heart is covered with various arteries such as Left Anterior Descending Artery (LAD), Left Circumflex Artery (LCX), and Right Coronary Artery (RCA) \cite{georgoulias2016frontiers, kim2011clinic}. Also, those arteries show various characteristic with CAO in ECG \cite{fuchs1982ecg}.

For conducting PCI, the physician needs to know the occlusion site of the coronary artery. Taking a CT scan may be more helpful for distinguishing the location, but there is not enough time for CT scan in an emergency situation. On the other hand, ECG can quickly measure the patient's condition, although only limited information is provided. Thus, we make the best use of ECG for localizing coronary occlusion site.

\section{Related work}
\label{sec:related}

One of the studies already shows the high performance at detecting coronary site location using decision tree \cite{gregg2014detect}. They deal with classifying LAD, LCX, and RCA, and their average sensitivity and specificity are 72\% and 92.5\% respectively using improved classification algorithm than their previous work \cite{gregg2012automated}. Sensitivity is a more important indicator than specificity, especially in emergency medicine. However, their sensitivity is 20.5\% lower than specificity. This means that only 7.5\% of normal patients are missed but CAO patients are missed 28\%.

The method for finding patients as much as possible is needed, and accordingly, various studies are already conducted. One of the studies suggests a preprocessing method for better performance in the same classifier \cite{park2019stemi}. That preprocessing method is dealt with in this research and several experiments for comparing and finding the better method.

\section{Proposed approach}
\label{sec:proposed}

In this section, we present our approach for localizing coronary occlusion site. We refer to the preprocessing method in the previous STEMI detection study \cite{park2019stemi}, and classification algorithm from study of Richard et al. \cite{gregg2014detect}. The above algorithm has not only simplified architecture but also improved performance than their previous work for achieving the same purpose \cite{gregg2012automated}. The preprocessing method is constructed with noise reduction and pulse extraction. The noise reduction is conducted via notch filter and high-pass filter. We apply the above preprocessing method before entering the data to the classifier. 

\begin{figure}[ht]
    \begin{center}
		\begin{tabular}{cc}
    			\includegraphics[width=0.60\linewidth]{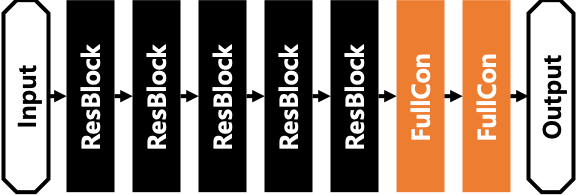} \vline & 
    			\includegraphics[width=0.20\linewidth]{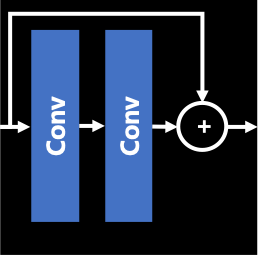}\\ 
		\end{tabular}
	\end{center}
	\vspace*{-5mm}
	\caption{In this figure the ResBlock, FullCon, and Conv means residual block, fully connected layer, and convolutional layer respectively. The residual block is constructed as the right subfigure.}
	\label{fig:archtecture}
\end{figure}

We construct the neural network architectures as shown in Figure~\ref{fig:archtecture} based on ResNet \cite{he2015deep}. We construct two CNN architecture using 1D convolutiona and 2D convolution separately. For comparing the performance, we use two kind of convolution, 1 Dimensional (1D) convolution and 2 Dimensional (2D) convolution. We use both architecture for experiment and then we will select more effective model.

\begin{figure}[ht]
    \begin{center}
		\includegraphics[width=0.30\linewidth]{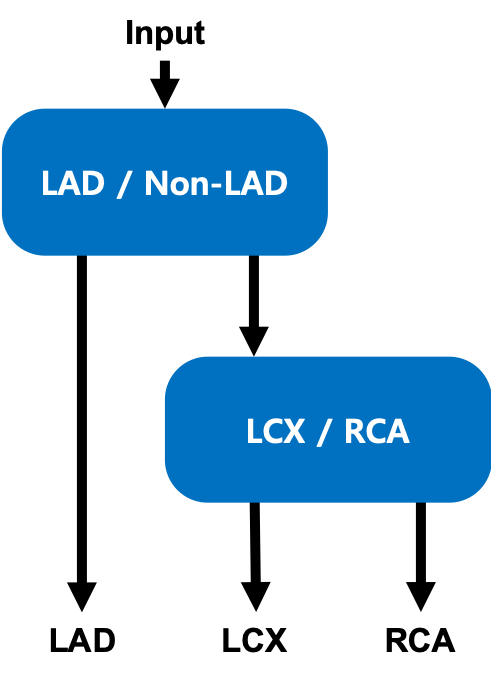}
	\end{center}
	\vspace*{-5mm}
	\caption{The CAO classification algorithm that constructed with two-stage. We name the classifier that classifies LAD or Non-LAD as stage-1 and the other, for classifying LCX or RCA, as stage-2.}
	\label{fig:classification}
\end{figure}

Then, we construct the CAO localization classification algorithm based on previous work \cite{gregg2014detect}. They used the three decision trees as a classifier. The first classifier is used for categorizing LAD or not, and the next two classifiers are used to classifying proximal LAD or non-proximal LAD and LCX or RCA respectively. However, we do not have the detailed label as proximal or non-proximal LAD, so we construct the classification algorithm as shown in Figure~\ref{fig:classification}.

Each stage of Figure~\ref{fig:classification} is constructed with CNN as shown in Figure~\ref{fig:archtecture}. We pursue to confirming the effectiveness of our preprocessing method based on denoising and pulse extraction at the above classification algorithm in this paper.

\section{Experiments}
\label{sec:experiments}

In this section, we present our dataset that originally provided by Seoul National University Bundang Hospital(SNUBH), and the experimental results. Firstly, we apply the preprocessing method as used in related work that method contains noise reduction and pulse extraction \cite{park2019stemi}. We call the each dataset as same as previous work. The two dataset, SNUBH-R and SNUBH-BP that as shown in Table~\ref{tab:dataset}, are used for experiments. The SNUBH-R contains raw ECG records, and the SNUBH-BP contains extracted pulse after applying both high-pass and notch filter. Then, we conduct the experiments using above dataset and the classifier as shown in Figure~\ref{fig:classification}.

\begin{table}[ht]
    \centering
    \caption{The ECG records are originally provided by SNUBH and it named SNUBH-R. The preprocessed dataset is named as SNUBH-BP.}
    \footnotesize
    \begin{tabular}{lrrr}
        \hline
            \textbf{Set} & \textbf{LAD} & \textbf{LCX} & \textbf{RCA} \\       
        \hline
            SNUBH-R & 419 & 70 & 283 \\
            SNUBH-BP & 4487 & 715 & 2591 \\
        \hline
    \end{tabular}
    \label{tab:dataset}
\end{table}

We use the accuracy, sensitivity, specificity, and Area Under the Receiver Operating characteristics Curve (AUROC) \cite{fawcett2006roc} as the performance indicator. The performance measurement is conducted for each of the two stages as the classifier consisted of two stages. The measured performances for each stage are summarized as Table~\ref{tab:performance1} and \ref{tab:performance2}.
\begin{table}[ht]
    \centering
    \caption{Performance of stage-1}
    \footnotesize
    \begin{tabular}{llllll}
        \hline
            \textbf{CNN} & \textbf{Set} & \textbf{Accuracy} & \textbf{Sensitivity} & \textbf{Specificity} & \textbf{AUROC} \\  
        \hline
            \multirow{2}{*}{1D-CNN}
            & SNUBH-R & 0.549 $\pm$ 0.039 & 0.453 $\pm$ 0.072 & 0.631 $\pm$ 0.064 & 0.560 $\pm$ 0.046 \\ 
            % & SNUBH-B & 0.760 $\pm$ 0.061 & 0.732 $\pm$ 0.151 & 0.783 $\pm$ 0.069 & 0.826 $\pm$ 0.054 \\ 
            % & SNUBH-RP & 0.906 $\pm$ 0.020 & 0.889 $\pm$ 0.026 & 0.919 $\pm$ 0.030 & 0.934 $\pm$ 0.021 \\ 
            & SNUBH-BP & 0.910 $\pm$ 0.020 & 0.895 $\pm$ 0.032 & 0.921 $\pm$ 0.024 & 0.932 $\pm$ 0.021 \\
        \hline
            \multirow{2}{*}{2D-CNN}
            & SNUBH-R & 0.569 $\pm$ 0.032 & 0.450 $\pm$ 0.168 & 0.658 $\pm$ 0.124 & 0.571 $\pm$ 0.042 \\ 
            % & SNUBH-B & 0.838 $\pm$ 0.114 & 0.815 $\pm$ 0.223 & 0.857 $\pm$ 0.168 & 0.862 $\pm$ 0.124 \\ 
            % & SNUBH-RP & 0.880 $\pm$ 0.061 & 0.838 $\pm$ 0.158 & 0.912 $\pm$ 0.029 & 0.887 $\pm$ 0.074 \\ 
            & SNUBH-BP & 0.915 $\pm$ 0.020 & 0.898 $\pm$ 0.029 & 0.927 $\pm$ 0.027 & 0.926 $\pm$ 0.020 \\
        \hline
    \end{tabular}\
    \label{tab:performance1}
\end{table}

When using the SNUBH-BP, dataset after preprocessing, the performance is highly improved than the case of using raw ECG record, SNUBH-R, both in the 1D-CNN and 2D CNN cases. The AUROC is increased 0.372 and 0.355 for 1D-CNN and 2D-CNN respectively. Moreover, It can be confirmed that the performance of 1D CNN is higher than 2D CNN.

\begin{table}[ht]
    \centering
    \caption{Performance of stage-2}
    \footnotesize
    \begin{tabular}{llllll}
        \hline
            \textbf{CNN} & \textbf{Set} & \textbf{Accuracy} & \textbf{Sensitivity} & \textbf{Specificity} & \textbf{AUROC} \\  
        \hline
            \multirow{2}{*}{1D-CNN}
            & SNUBH-R & 0.770 $\pm$ 0.034 & 0.944 $\pm$ 0.037 & 0.062 $\pm$ 0.058 & 0.513 $\pm$ 0.082 \\ 
            % & SNUBH-B & 0.779 $\pm$ 0.023 & 0.958 $\pm$ 0.030 & 0.050 $\pm$ 0.064 & 0.491 $\pm$ 0.062 \\ 
            % & SNUBH-RP & 0.741 $\pm$ 0.028 & 0.882 $\pm$ 0.033 & 0.221 $\pm$ 0.091 & 0.588 $\pm$ 0.068 \\ 
            & SNUBH-BP & 0.775 $\pm$ 0.030 & 0.898 $\pm$ 0.028 & 0.319 $\pm$ 0.103 & 0.663 $\pm$ 0.050 \\
        \hline
            \multirow{2}{*}{2D-CNN}
            & SNUBH-R & 0.715 $\pm$ 0.044 & 0.854 $\pm$ 0.059 & 0.150 $\pm$ 0.075 & 0.518 $\pm$ 0.058 \\ 
            % & SNUBH-B & 0.721 $\pm$ 0.065 & 0.865 $\pm$ 0.096 & 0.133 $\pm$ 0.110 & 0.503 $\pm$ 0.067 \\ 
            % & SNUBH-RP & 0.718 $\pm$ 0.032 & 0.870 $\pm$ 0.043 & 0.169 $\pm$ 0.084 & 0.522 $\pm$ 0.047 \\ 
            & SNUBH-BP & 0.764 $\pm$ 0.038 & 0.892 $\pm$ 0.050 & 0.284 $\pm$ 0.117 & 0.612 $\pm$ 0.062 \\
        \hline
    \end{tabular}\
    \label{tab:performance2}
\end{table}

In stage-2, the performance is also improved with SNUBH-BP set. Thus, we confirm that our preprocessing technique can help to improve the CAO classification ability. However, the above two tables show that the 1D-CNN shows better performance than 2D-CNN, so if who wants to use CNN for classifying CAO the 1D-CNN is recommended.

\section{Conclusion}
\label{sec:conclusion}

We do not conduct the experiment for finding the optimal structure of the CNN and the hyperparameters for CNN. However, our experiments show that our preprocessing technique, based on both noise reduction and pulse extraction, can improve the classification performance with simple CNN. The performance is improved as much as 0.372 and 0.150 in stage-1 and stage-2 respectively with 1D-CNN. The CAO classification performance will be highly improved if the optimal CNN structure is found with fine hyperparameter. However, when using our preprocessing method, everyone can achieve high performance without finding the fine classifier.

\section*{Acknowledgement}
\label{sec:acknowledgement}

This research was supported by grant No.~2017R1A2B4004503 and Hankuk University of Foreign Studies Research Fund. Also, this research was in line with the Declaration of Helsinki and approved by the Seoul National University Bundang Hospital institutional review board in 2019 (B-1811-502-003).

\bibliographystyle{unsrt}  
\bibliography{references}  %%% Remove comment to use the external .bib file (using bibtex).


% George Kour and Raid Saabne.
% \newblock Real-time segmentation of on-line handwritten arabic script.
% \newblock In {\em Frontiers in Handwriting Recognition (ICFHR), 2014 14th
%   International Conference on}, pages 417--422. IEEE, 2014.

% George Kour and Raid Saabne.
% \newblock Fast classification of handwritten on-line arabic characters.
% \newblock In {\em Soft Computing and Pattern Recognition (SoCPaR), 2014 6th
%   International Conference of}, pages 312--318. IEEE, 2014.

% \bibitem{hadash2018estimate}
% Guy Hadash, Einat Kermany, Boaz Carmeli, Ofer Lavi, George Kour, and Alon
% \newblock Estimate and replace: A novel approach to integrating deep neural
%   networks with existing applications.
% \newblock {\em arXiv preprint arXiv:1804.09028}, 2018.

% \end{thebibliography}

\end{document}